# Superconducting Properties and Phase Diagram of Indirectly Electron-Doped $(Sr_{1-x}La_x)Fe_2As_2$ Epitaxial Films Grown by Pulsed Laser Deposition

Hidenori Hiramatsu, Takayoshi Katase, Toshio Kamiya, and Hideo Hosono

*Abstract*— A non-equilibrium phase $(Sr_{1-x}La_x)Fe_2As_2$ was formed by epitaxial film-growth. The resulting films emerged superconductivity along with suppression of the resistivity anomaly that is associated with magnetic and structural phase transitions. The maximum critical temperature was 20.8 K, which is almost the same as that of directly electron-doped $Sr(Fe_{1-x}Co_x)_2As_2$. Its electronic phase diagram is much similar to that of $Sr(Fe_{1-x}Co_x)_2As_2$, indicating that the difference in the electron doping sites does not influence the superconducting properties of 122-type $SrFe_2As_2$.

## I. Introduction

The discovery of an Fe-based layered superconductor, 1111-type $LaFeAs(O_{1-x}F_x)$ [1] revived intensive researches on high critical temperature ($T_c$) superconductors to find new superconductors. To induce high-$T_c$ superconductivity, carrier doping with appropriate aliovalent dopants has been examined. Maximum $T_c$ values for different Fe-based parent materials have been obtained by 'indirect doping' [2]–[4], in which the dopant replaces a site other than the Fe site, because the direct doping at the Fe site gives a serious perturbation to the electronic structure of the Fermi surface and the carrier transport.

122-type $AEFe_2As_2$ ($AE$ = alkaline earth such as Ba, Sr, and Ca) is one of the parent materials of the Fe-based high-$T_c$ superconductors [4]. Recently, indirect rare-earth ($RE$) doping to $AEFe_2As_2$ was achieved using a high-pressure synthesis process for $(Sr_{1-x}La_x)Fe_2As_2$ polycrystals [5], a melt-growth process using a flux for $(Ca_{1-x}RE_x)Fe_2As_2$ ($RE$ = La – Nd) single crystals [6]–[8], and a non-equilibrium thin-film growth process for $(Ba_{1-x}RE_x)Fe_2As_2$ ($RE$ = La – Nd) [9] [10]. Among them, it is noteworthy that the Pr-doped $CaFe_2As_2$ $((Ca_{1-x}Pr_x)Fe_2As_2)$ single crystal exhibits maximum $T_c$ of 49 K [7], which is the highest $T_c$ in the 122-type $AEFe_2As_2$ compounds. In addition, even though the $CaFe_2As_2$ doped with the largest size $RE$ (La), the maximum $T_c$ is as high as 42.7 K [8]. However, the origin of its high-$T_c$ still remains controversy because its shielding volume fraction is as low as < 1 %.

The maximum $T_c$ of the indirectly electron-doped $(Sr_{1-x}La_x)Fe_2As_2$ polycrystals fabricated under a high-pressure is 22 K [5], which is slightly higher but very close to that of the directly electron-doped $Sr(Fe_{1-x}Co_x)_2As_2$ (20 K) [11] and is much lower than that (42.7 K) of the $(Ca_{1-x}La_x)Fe_2As_2$ single crystals [8]. The $T_c$ of the $(Sr_{1-x}La_x)Fe_2As_2$ polycrystals were almost independent of the La concentration due to inhomogeneous replacement of the La dopants at the Sr sites. However, in the case of the $(Ba_{1-x}RE_x)Fe_2As_2$ epitaxial films ($RE$ = La and Ce) [9] [10], we have succeeded in homogeneous doping of the $RE$ dopants in the films by non-equilibrium pulsed laser deposition (PLD) and observing a superconducting dome from under-doped to completely over-doped (i.e., suppression of superconductivity) regions irrespective of the largely different ion radii of $Ba^{2+}$ (142 pm) and $RE^{3+}$ dopants (116 pm for La and 114 pm for Ce). Therefore, we expected that the PLD process could effectively stabilize the $RE$ dopants in other 122-type epitaxial films composed of a smaller $AE$ such as $SrFe_2As_2$ (126 pm for Sr) and $CaFe_2As_2$ (112 pm for Ca); i.e., this expectation would lead to higher $T_c$ close to those of the $(Ca_{1-x}RE_x)Fe_2As_2$ crystals, based on that the difference in ion radii between the $AE$ and the $RE$ dopants becomes smaller than that of the $AE$ = Ba case and the smallest $AE$ = Ca 122-type material exhibits the highest $T_c$. However, very recently, it has turned out to be difficult to obtain $CaFe_2As_2$ epitaxial films by PLD [12].

Thus, in this study, we selected the $SrFe_2As_2$ system and succeeded in indirect electron-doping of La by PLD. The superconducting properties of resulting $(Sr_{1-x}La_x)Fe_2As_2$ epitaxial films are discussed in comparison with the results of directly electron-doped $Sr(Fe_{1-x}Co_x)_2As_2$.

## II. Experimental

### A. Thin Film Growth

Thin films of $(Sr_{1-x}La_x)Fe_2As_2$ (thickness: ~200 nm) were grown on (La, Sr)(Al, Ta)$O_3$ (LSAT) (001) single-crystal substrates with a PLD system using a second harmonic of a pulsed Nd:YAG laser as an excitation source [13]. La-added $SrFe_2As_2$ sintered disks with nominal $x = 0 - 0.25$, which were segregated mixtures of undoped $SrFe_2As_2$ and LaAs, were employed as PLD targets because the substitution of La dopants at Sr sites is not stable at the thermal equilibrium and has not been realized by conventional solid-state reactions [5]. Detailed procedures of the target syntheses by the solid-state reaction were reported in [9] [10]. The film-growth temperature was set at ~750 °C, which was optimized for epitaxial growth of $SrFe_2As_2$ [12]–[14]. The other film-growth conditions were same as the optimized ones for epitaxial growth of $Ba(Fe_{1-x}Co_x)_2As_2$ [12].

### B. Characterization

Doped La concentrations in the films ($x_{film}$: atomic ratios of La / (Sr+La)), varied by the nominal $x$ in the PLD targets, were determined with an electron-probe microanalyzer (EPMA). Because we used LSAT single crystals as substrates, the La and Sr in the substrates interfere the quantitative chemical analyses of the films. This effect is serious and it was difficult to separate the Sr L$\alpha$ (= 1.8 keV) signals from the films and the substrates if the electron beam acceleration at 6 keV perpendicular to the film surface was employed. Therefore, based on simulated results of probing depth profiles, the samples were tilted at 70 degrees in order to safely detect the Sr L$\alpha$ signal only from the film bulk region. The $x_{film}$ were determined using the tilted sample-configuration. Since the La L$\alpha$ (= 4.7 keV) probing-depth is not deep (~30 nm when 6 keV), the mapping image of La was measured to examine its homogeneity. Crystalline phases and lattice parameters were examined by x-ray diffraction (XRD, radiation: CuK$\alpha_1$) at room temperature. $c$- and $a$-axes lattice parameters were determined by out-of-plane and in-plane measurements, respectively [15]. Temperature dependences of resistivity ($\rho$–$T$) were measured using a physical property measurement system from 2 to 300 K. Transport critical current densities ($J_c^{trans}$) were determined with a criterion of 1 $\mu$V/cm from current density – voltage curves using 1-mm-long and 2.3-mm-wide bridges. Magnetization hysteresis loops, from which the magnetic $J_c$ ($J_c^{mag}$) were extracted, and temperature dependence of magnetic susceptibilities ($\chi$–$T$) were measured with a vibrating sample magnetometer.

### III. RESULT AND DISCUSSION

Figure 1 (a) shows relationship between $x_{film}$ and nominal $x$. Each $x_{film}$ is roughly two times larger than that of the nominal $x$, but is controlled almost linearly by the nominal $x$. As shown in Fig. 2(a), sharp 00$l$ diffractions of the $(Sr_{1-x}La_x)Fe_2As_2$ phases were observed from the low $x_{film}$ = 0.18 to the highest $x_{film}$ = 0.54, along with the diffraction from an impurity Fe phase. They shifted to higher angles from the diffraction angles of the undoped $SrFe_2As_2$ epitaxial film with increase in $x_{film}$. However, the clear segregation of the impurity phase of LaAs was observed at the highest $x_{film}$ of 0.54 as indicated in the top panel of Fig. 2(a) by an arrow at $2\theta$ = 29.08 degrees. We confirmed by XRD that the obtained $(Sr_{1-x}La_x)Fe_2As_2$ films with $x_{film} \leq 0.48$ were grown heteroepitaxially on the LSAT substrates with the epitaxial relationship of (001) $(Sr_{1-x}La_x)Fe_2As_2$ || (001) LSAT for out-of-plane and [100] $(Sr_{1-x}La_x)Fe_2As_2$ || [100] LSAT for in-plane. We also examined

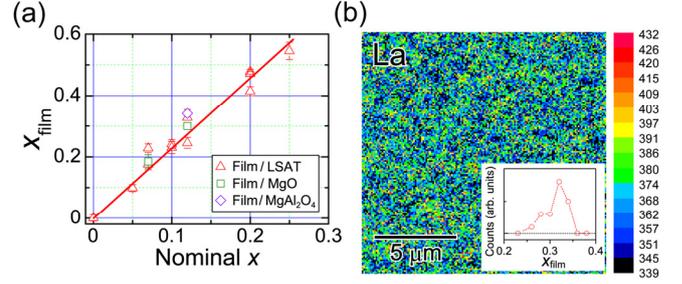

Fig. 1. Chemical composition. (a) La concentration in the film $x_{film}$ measured by EPMA as a function of nominal $x$ in the PLD target. Triangles, squares, and a diamond indicate the cases of LSAT, MgO, and MgAl$_2$O$_4$ substrates, respectively. (b) Composition mapping image of La for the $(Sr_{0.68}La_{0.32})Fe_2As_2$ epitaxial film on LSAT substrate. The numbers on right of the scale bar indicate the signal intensity in counts per pixel. The inset of (b) shows the relationship between counts and $x_{film}$.

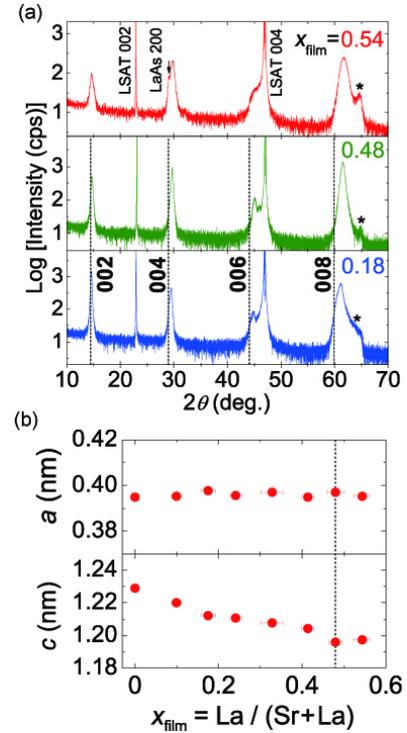

Fig. 2. (a) Out-of-plane XRD patterns of the $(Sr_{1-x}La_x)Fe_2As_2$ films with $x_{film}$ = 0.18, 0.48, and 0.54. The asterisks indicate the diffraction peaks from Fe impurity. Dotted lines are the peak positions of the 00$l$ diffractions of the undoped $SrFe_2As_2$ epitaxial film. (b) $x_{film}$ dependence of lattice parameters $a$- and $c$-axes. The dotted line shows the solubility limit of La dopant in the $(Sr_{1-x}La_x)Fe_2As_2$ epitaxial film.

MgO and MgAl$_2$O$_4$ (001) single-crystal substrates, the $x_{film}$ were close to those of LSAT case (Fig. 1(a)), but the 110 diffraction of the $(Sr_{1-x}La_x)Fe_2As_2$ phase were always observed in the out-of-plane XRD patterns; i.e., heteroepitaxial growth could not be observed. This is probably due to the in-plane lattice mismatches between $SrFe_2As_2$ and the substrates; LSAT has the smallest mismatch (–1.5 %) among these three kinds of substrates (+6.8 % for MgO and +2.9 % for MgAl$_2$O$_4$). Figure 2(b) shows the lattice parameters of the films as a function of $x_{film}$. The $a$-axis length was independent of the $x_{film}$, but a systematical shrinkage of the $c$-axis length was observed as $x_{film}$ increased from 0 to 0.48. EPMA mapping image (Fig. 1(b)) also proves that the La dopant is distributed homogeneously in the film (within 10 %). In contrast, no change in $c$-axis length was observed from $x_{film}$ = 0.48 to 0.54. Therefore, we evaluate that the solubility limit of La dopant is $x_{film}$ = 0.48.

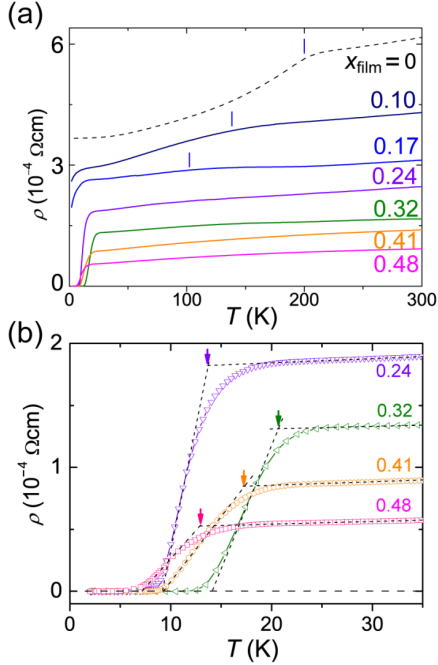

Fig. 3. (a) $\rho$–$T$ curves of the $(Sr_{1-x}La_x)Fe_2As_2$ epitaxial films with $x_{film}$ = 0 – 0.48. The vertical lines indicate positions of the resistivity anomaly. (b) Enlarged view of the superconducting transition of the films with $x_{film}$ = 0.24 – 0.48.

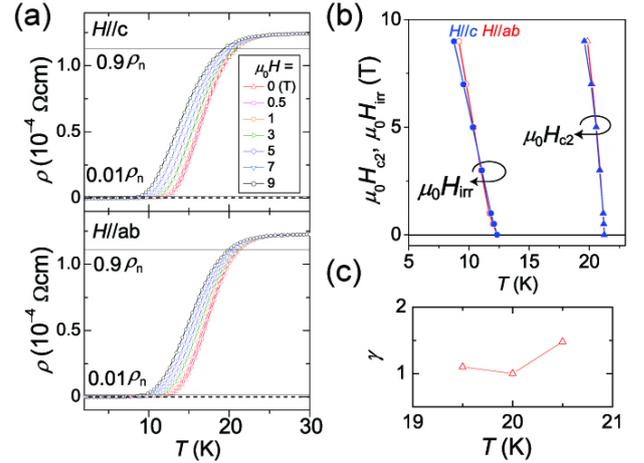

Fig. 4. (a) Magnetic field ($H$) dependence of $\rho$–$T$ curves for the optimally doped $(Sr_{0.68}La_{0.32})Fe_2As_2$ epitaxial film up to 9 T. Upper and lower panels show that the fields are applied parallel to the $c$-axis and to the $ab$-plane, respectively. (b) Temperature dependences of upper critical field ($H_{c2}$) and irreversibility field ($H_{irr}$). (c) Estimated anisotropy factor $\gamma = H_{c2}^{//ab} / H_{c2}^{//c}$.

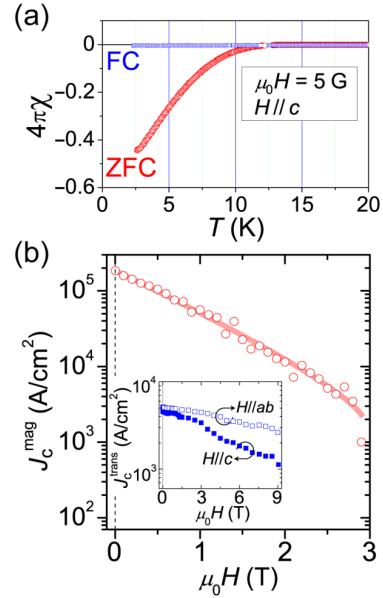

Fig. 5. Magnetic properties of the optimally doped $(Sr_{0.68}La_{0.32})Fe_2As_2$ epitaxial film. (a) $\chi$–$T$ curves measured by zero-field cooling (ZFC) and during field cooling (FC) modes. (b) Magnetic field dependences of $J_c^{mag}$ at 2 K. Inset shows $J_c^{trans}$ of the film at 2 K. The closed and open symbols in the inset indicate that the fields are applied parallel to the $c$-axis and the $ab$-plane, respectively.

The dotted line in Fig. 3(a) shows the $\rho$–$T$ curve of an undoped $SrFe_2As_2$ epitaxial film, which was measured immediately after taken out from the PLD chamber, because the undoped $SrFe_2As_2$ epitaxial film is sensitive to water vapor in air [16]. A resistivity anomaly corresponding to a structural phase transition and an antiferromagnetic (AFM) ordering is observed at $T_{anom}$ = 200 K, which is roughly same as those reported previously for undoped bulk samples [17]. With increase in $x_{film}$, normal state $\rho$ values at 300 K gradually decreased and the $T_{anom}$ shifted to lower temperature. A $T_c^{onset}$ began observed on the film with $x_{film}$ = 0.1, although the resistivity anomaly is still observed at $T_{anom}$ = 135 K. This result implies that superconductivity and AFM order coexist at the low doping level. The coexistence was observed also at $x_{film}$ = 0.17. With further increase in $x_{film}$, the resistivity anomaly was not detected and $T_c^{zero}$ appears at $x_{film}$ = 0.24 (Fig. 3(b)). The maximum $T_c^{onset}$ was 20.8 K at $x_{film}$ = 0.32. The $\Delta T_c$ (=$T_c^{onset} - T_c^{zero}$) of the film was as large as 6.6 K, which is comparable to that of $Sr(Fe_{1-x}Co_x)_2As_2$ epitaxial film [14]. After that, $T_c^{onset}$ decreased to 12.9 K as $x_{film}$ further increased to 0.48, which is the solubility limit of La dopant. Due to the limitation, a completely over-doped region, where superconductivity should disappear, was not observed.

Figure 4(a) shows the effects of external magnetic fields ($H$) on the superconducting transitions of the optimally doped $(Sr_{1-x}La_x)Fe_2As_2$ epitaxial films with $x_{film}$ = 0.32. The $H$ were applied parallel to the $c$-axis (upper) and to the $ab$-plane (lower) of the film. It was observed that $T_c^{onset}$ and $T_c^{zero}$ decreased and $\Delta T_c$ increased with increasing $H$. From the results in Fig. 4(a), upper critical magnetic fields ($H_{c2}$) and irreversibility fields ($H_{irr}$) were respectively estimated using criterions of $0.9\rho_n$ and $0.01\rho_n$, where $\rho_n$ is the $\rho$ at 30 K (Fig. 4(b)). The values of $H_{c2}(0)$ were obtained using the Werthamer–Helfand–Hohenberg model by $H_{c2}(0) = -0.693T_c \times (dH_{c2}/dT)_{T=T_c}$ [18]. The estimated $H_{c2}^{//ab}(0)$ and $H_{c2}^{//c}(0)$ were 131 and 95 T, respectively. The corresponding coherence lengths $\xi_{ab}(0)$ and $\xi_c(0)$, estimated by the Ginzburg-Landau relation, were 1.8 and 1.3 nm, respectively. The anisotropy factors $\gamma = H_{c2}^{//ab} / H_{c2}^{//c}$ were estimated to be 1 – 1.5 (Fig. 4(c)), substantiating its small anisotropy. All these physical parameters obtained are comparable to those of other 122-type compounds [19] [20].

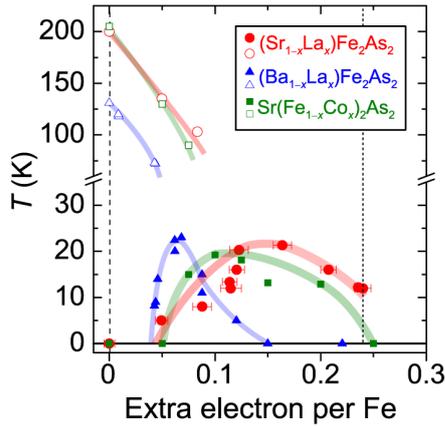

Fig. 6. Electronic phase diagram of $(Sr_{1-x}La_x)Fe_2As_2$ epitaxial films. The $T_{anom}$ and $T_c^{onset}$ are indicated by open and closed symbols, respectively. Those of $Sr(Fe_{1-x}Co_x)_2As_2$ and $(Ba_{1-x}La_x)Fe_2As_2$ are shown for comparison.

Figure 5(a) shows $\chi - T$ curves of the optimally doped $(Sr_{0.68}La_{0.32})Fe_2As_2$ epitaxial film. Clear diamagnetic signal starting at 12 K was observed. The shielding volume fraction was estimated to be > 40 % at 2 K, proving the bulk superconductivity of the film. The $J_c^{mag}$ at 2 K was 0.2 MA/cm$^2$ but it decreased exponentially as $H$ increased (Fig. 5(b)). The $J_c^{trans}$ was much lower than $J_c^{mag}$ as seen in inset of Fig. 5(b). The underestimation of the $J_c^{trans}$ is similar to those of $Sr(Fe_{1-x}Co_x)_2As_2$ films [21] [22], implying that the $(Sr_{1-x}La_x)Fe_2As_2$ films also have a granularity problem (weak link). However, $J_c^{trans}$ //ab were larger than those of $J_c^{trans}$ //c. This result suggests that the pinning mechanism, which may be similar to that of a $Ba(Fe_{1-x}Co_x)_2As_2$ film [23], is different from each other because the c-axis pinning was clearly observed in the case of $Sr(Fe_{1-x}Co_x)_2As_2$ epitaxial films [21].

Figure 6 shows the electronic phase diagram of the $(Sr_{1-x}La_x)Fe_2As_2$ epitaxial films. The doping concentrations are normalized as the doped carriers per Fe (i.e., $x_{film}/2$) to compare with the phase diagram of the directly electron-doped $Sr(Fe_{1-x}Co_x)_2As_2$ [11]. That of the $(Ba_{1-x}La_x)Fe_2As_2$ epitaxial films is also shown for comparison [9]. The AFM transition at $T_{anom}$ is suppressed as $x_{film}$ increased, and the suppression rate of $T_{anom}$ (i.e., $|dT_{anom}/dx_{film}|$) is almost the same for each dopant and $(Ba_{1-x}La_x)Fe_2As_2$. Coexistence of superconductivity and an AFM state is stabilized only in the low doping region. The maximum $T_c^{onset}$ for the $(Sr_{1-x}La_x)Fe_2As_2$ films is 20.8 K, which is almost the same as those of $Sr(Fe_{1-x}Co_x)_2As_2$ (20 K) [11] and $(Ba_{1-x}La_x)Fe_2As_2$ (22.4 K) [9] but much lower than that of $(Ca_{1-x}La_x)Fe_2As_2$ (42.7 K) [8]. It should be noted that the maximum $T_c^{onset}$ was achieved at $x_{film}/2 = 0.16$, which is also almost the same as that for the directly doped $Sr(Fe_{1-x}Co_x)_2As_2$; i.e., the superconducting dome shape and coexistence of AFM and superconductivity are qualitatively similar to those of the $Sr(Fe_{1-x}Co_x)_2As_2$ although the superconducting dome of the $(Sr_{1-x}La_x)Fe_2As_2$ epitaxial films did not reach a completely over-doped region (i.e., suppression of superconductivity) because of the lower solubility limit of La dopant. These results, similar to $(Ba_{1-x}La_x)Fe_2As_2$ [9], sharply contrast with those of 1111-type iron pnictides, in which maximum $T_c$ (55 K) of indirect-doped $SmFeAs(O_{1-x}F_x)$ [2] and $SmFeAs(O_{1-x}H_x)$ [3] is much higher than that of direct-doped $Sm(Fe_{1-x}Co_x)AsO$ (17 K) [24]. The identical effect of the indirect and the direct electron-dopings of $SrFe_2As_2$ on the superconductivity is similar to the case of $BaFe_2As_2$ [9], indicating a common nature for the 122-type electron-doped $AEFe_2As_2$ ($AE$ = Sr and Ba) superconductors. There are two plausible reasons for explaining this observation. First is due to the pairing symmetry. Because it has been theoretically proposed that a non-sign-reversal $s_{++}$ symmetry is robust to the direct doping while a sign reversal $s_{+-}$ is sensitive [25], the above observation is more consistent with a $s_{++}$ wave state, not a $s_{+-}$ wave state. Second is dimensionality of electronic structure of the 122-type $AEFe_2As_2$. Although the $AEFe_2As_2$ has a layered structure, its band structure has a distinct dispersion also along the Γ–Z direction [26] [27]. This fact means that electronic state of the FeAs layer cannot be separated from that of the $AE$ layer. The obtained result may also be reasonably understood from this view point. Two-dimensional nature in electronic state is prominent for the 1111-type material. It is thus reasonable that marked effect of doping mode on $T_c$ in the 1111-type iron pnictides.

## IV. SUMMARY

Indirect electron-doping to a 122-type iron pnictide, $SrFe_2As_2$, was achieved by employing a non-equilibrium PLD process. The maximum $T_c$ was 20.8 K, which was the same as that of directly electron-doped $Sr(Fe_{1-x}Co_x)_2As_2$ (20 K) and was much lower than that of indirectly electron-doped $(Ca_{1-x}La_x)Fe_2As_2$ (42.7 K). We found that indirect-doped $(Sr_{1-x}La_x)Fe_2As_2$ and direct-doped $Sr(Fe_{1-x}Co_x)_2As_2$ exhibit the identical relationship between $T_c$ and doping levels, which is similar to that of $BaFe_2As_2$. This finding indicates that the primary factor controlling $T_c$ in $SrFe_2As_2$ is not a doping mode; i.e., indirect or direct, but the polarity and concentration of doped carriers.

ACKNOWLEDGMENT

This work was supported by the Japan Society for the Promotion of Science (JSPS), Japan, through the Funding Program for World-Leading Innovative R&D on Science and Technology (FIRST Program). H. Hiramatsu acknowledges a support from the Asahi Glass Foundation.